# FISLAB - the Fuzzy Inference Tool-box for SCILAB

**Univ.Assist. Simona Apostol**
"Tibiscus" University of Timisoara, Romania

**Rezumat:** Lucrarea de fata reprezinta "Pachetul de programe Fislab destinat dezvoltarii regulatoarelor fuzzy in cadrul mediului Scilab", in care sunt prezentate cateva aspecte generale,cerinte de utilizare cat si modul de lucru al mediului Fislab.
In partea a doua a articolului sunt descrise cateva functii Scilab din cadrul pachetului Fislab.

## 1. General aspects

**Fislab** ([Ort97]) is a fuzzy inference systems tool-box for SCILAB.FISLAB which allows different arithmetic operators, fuzzification and defuzzifications algorithms, the implication relations, and different methods of approximate reasoning such as Compositional Rule of Inference (CRI) and Approximate Analogical Reasoning Scheme based on Similarity Measure. FISLAB is based on a MATLAB inference tool-box developed by Prof. Zadeh called FISMAT.

Fislab contains a set with specific functions for:
- Linguistic variables definition and afferent linguistics terms through the ownership functions.
- Fuzzification in firm information;
- Definition of base rules and of inference method;
- Effectuation of defuzzification.

According to [Ort97], in the development of the Fislab software package it was started from a similar package, Fismat, advanced by A.Lotfi.

## 2. Utilization requests

In order to run FISLAB the following software is needed: Scilab version 2.2 or a more recent one. The software has been tested in a PC running Linux, and in a SPARC Station 5 running SunOS. After the installation of Scilab medium, in the sub-folder *contrib* of *scilab-4.0* folder, the folder Fislab .tar.gz (it is specified xfFislab.tar if WinZip*8.1* is not used) can be extracted with WinZip. Start running Scilab medium (scilab-4.0/bin/runscilab.exe) and then write the next two commands lines (which specify the path): *exec('c:scilab\scilab-4.0\contrib\FISLAB\builder.sce')* and *exec('c:scilab\scilab-4.0\contrib\FISLAB\loader.sce')*. At each starting of Scilab, if the use of Fislab is needed, we must input the two specified lines. From this moment Fislab software can be used.

## 3. The Fislab Application

Fislab contains a collection of Scilab functions belonging to the next modules specified for a fuzzy regulator ([PP97]), with the diagram in fig.4.3:
- Fuzzifications subsystem, which contains linguistic variables definition and afferent terms through the ownership functions;
- Inference diagram;
- Base rules;
- A defuzzifications subsystem.

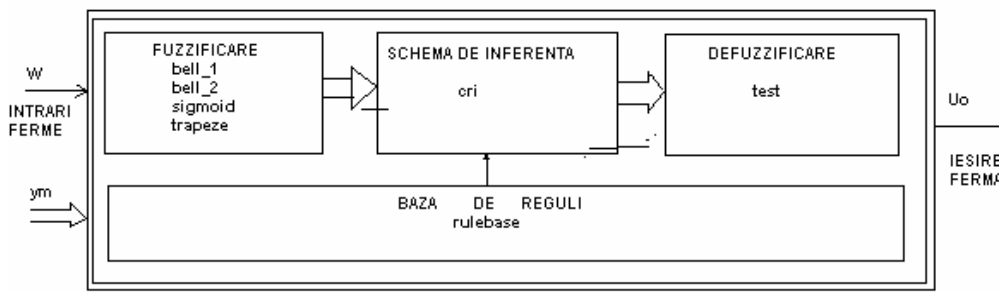

*Fig. 1. Fuzzy regulator diagram with afferents functions by Fislab package*

The fuzzyfication module contains the next Fislab functions:
- *bell_1;*
- *bell_2;*
- *sigmoid ;*
- *trapeze.*

Compositional Rule of Inference is represented with the next Fislab function:
- cri.

The rules base contains the next Fislab function:
- rulebase.

Defuzzyfication subsystem contains the next Fislab function:
- test.

That function is in the *test.sci* folder.

Comparing to Fismat, the major differences are ([Ort97]) in the output representation subsystem and in Fislab defuzzyfication subsystem.

Note that the Fislab can permit the development of a fuzzy regulators class, meaning **the development of the fuzzy regulators (RG-F) of the Mamdani type**. In order to emphasize what characterizes these regulators, it should be reminded that generally the rules diagram in **RG-F case of Takagi-Sugeno type** is represented in relation (1) ([Dra01]), for the $R_i$ rule of fuzzy regulation from the rules base:

$$R_i : \bigwedge_{k=1}^{m}(X_k = A_{k,i_k}) \Rightarrow C = f_i(X_1,...,X_m) \: pentru \: i \in \{1,...,n\}, i_k \in \{1,...,\alpha_k\} \quad (1)$$

where *fi* represents an application (particularly, a linear one) from the multiplication space $X_1 \times ... \times A_m$ of the input variables $X_1,...,A_m$ to $C$ the basic set of the $C$ output variables (from the conclusion), and $A_{k,i_k}$ represents the linguistics terms of the input linguistic variables

Note that for the fuzzy regulator rules (1) there is not aprioristic to define the $B_i$ linguistic terms from the conclusion and that these rules do not contain ponder factors. RG-F of the Takagi-Surgeon type provides an F(x) static input-output feature, afferent to the regulator and true for the entire definition domain, made up of more $i_f$ functions, true for domains/parts.

107

The constant connection between RG-F of Takagi-Sugeno type and the RG-F of Mamdani type can be easily illustrated by replacing in the relation (4.3.1) the $f_i$ functions from the conclusion by $c_i^0$ singletons ($c_i^0$ are the modal values of the singletons). The immediate regulations the fuzzy regulation rules of the RG-F of Mamdani type can be declare like below general form (2):

$$R_i : \bigwedge_{k=1}^{m} \left( X_k = A_{k,i_k} \right) \Rightarrow C = c_i^0 \ pentru \ i \in \{1,...,n\}, i_k \in \{1,...,\alpha_k\} \qquad (2)$$

where the singleton from conclusion of the rule can be replaced by any linguistic term of the linguistic output variable.

For details concerning fuzzy regulator the following study can be consulted ([PP97]).

## 4. The description of the Scilab Functions within the Fislab package

Following diagram of the Fislab medium in 1.0 version according to (Fislab, 1997) the next Scilab functions are available.

*Note:* Folder *intrdemo.sci* represents a demonstrative folder for Fislab. The folder *intrdemo* – which is not, in fact, a function – must be executed when the afferent Fislab demo is annulated.

**Scilab functions used for the definition of ownership functions**
As part of this paragraph there are presented main Scilab functions by which different types of belonging functions can be defined which are available to the user in FLT-Scilab mode.

**Function bell_1**
The belonging function of Gauss **bell_1** type form Fislab, correspond to the Matlab:**gaussmf function.**
**File:** *bell_1.sci*
**Description:** It is a function which implements the belonging function of Gauss' bell type with three parameters: *a, b, c*. It returns an Y matrix with the same dimension like X matrix; each element from Y is an appartenention grad.

The function implements the next expression:

Y=exp(-((((X -c)/a)^2)^b) )
Fislab syntax function for the belonging function is:
y=bell_1(x,param), param=a b c.
An example of utilization of this function is presented next:
-->x=(0:0.1:10)';
-->y=bell_1(x,-1,3,4);
-->plot2d(x,y)
The result of the development of this example is illustrated in fig. 2.

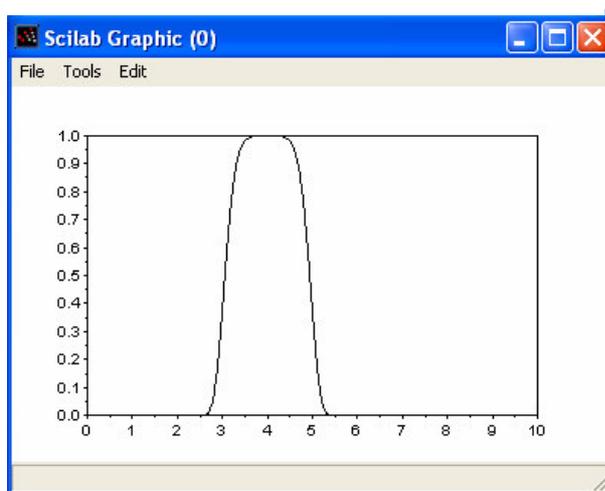

*Fig. 2. Example for bell_1 function*

**Function bell_2**
The generalized bell belonging function **bell_2** from Fislab, correspond to Matlab: **gbellmf** function
**File:** *bell_2.sci*
**Description:** It is a function of bell-shaped class generalized with three parameters a,b,c..It returns a Y matrix by the same dimension like X. The elements of X are firm values from the associated speech universe associated a with fuzzy set with a given form Ap function; the Y elements are belonging grades (see previous function).
The function implements the next formula:
Y=1/(1+(((X - c)/a)^2)^b)
where c is the centered value, usually b>0, a and b characterize the form of the bell.
Fislab function syntax corresponding to this belonging function is:
y=bell_2(x,param), param=a b c.
Next is illustrated the utilization for this function.

```
-->x=(0:0.1:10)';
-->y=bell_2(x,-1,3,4);
-->plot2d(x,y)
```
The results for this example are displayed in fig. 3.

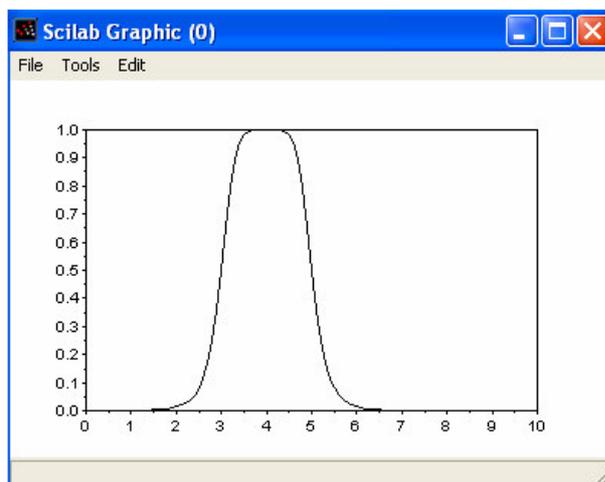

*Fig. 3. Example of bell_2 usage*

**Sigmoid function**
Sigmoid belonging function of the Fislab, corresponds to Matlab:**sigmf** function**.**
**File:** *sigmoid.sci*
**Description:** This function implements in Scilab an belonging function of sigmoid type with two a and c, parameters. The function returns a Y matrix with the same dimension as X; each element of Y is an belonging grad. That function has implicit values for a and c, 1 respectively 0.
The belonging function of sigmoid type has the next analytic expression:

$$\mu(x) = \frac{1}{1+e^{-a(x-c)}}$$

where for a>0 belonging function is opened to the right, and for a<0 to the left. The afferent function syntax for Fislab is:
   y=sigmoid(x,param),  param=a,c.
   An example of using the function results by developing the next sequence of Scilab program:
```
-->x=(0:0.1:10)';
-->y=sigmoid(x,1,5);
-->plot(x,y)
```
110

The result for this example is graphically represented in fig. 4.

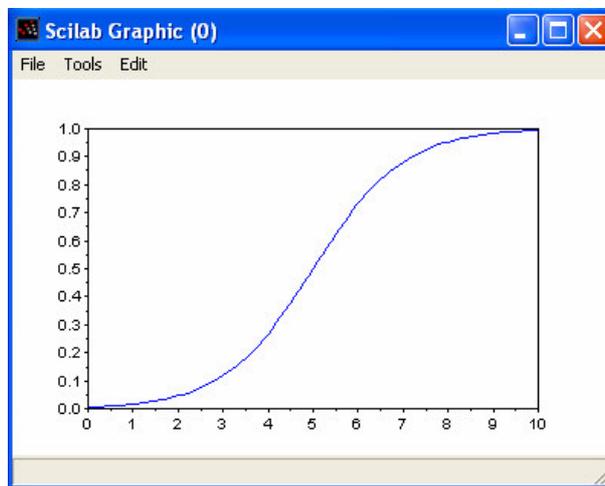

*Fig. 4. Sigmoid function utilization*

**Trapeze function**
The belonging function of trapezoidal **trapeze** type from Fislab corresponds in Matlab to: **trapmf** ([PP97]).
**File:** *trapeze.sci*
**Description:** This function implements the belonging functions of trapezoidal type which correspond to an X matrix which contains firm values. The functions having the parameters a, b, c which are the ramp, the slides, and the center of trapeze (supposed isosceles). This function has as implicit parameters for b and c, 0, respectively 1, for a.

The syntax for Fislab function afferent to the belonging function of trapezoidal type is described by the next expression:

y = trapeze(x, param),

in which:

param =a,b,c.

An example of utilization is represented in the next program sequence:
-->x=(0:0.1:10)';
-->[y]=trapeze(x,1,0,0);
-->plot2d(x,y)

The result of the developing example is represented in fig. 5.

In the next example a belonging function of triangular type is obtained, efficiently used at TL afferent to the linguistic variables of the fuzzy regulators (RG-F), with the result illustrated in fig. 6:

111

```
-->x=(-10:5:10)';
-->[y]=trapeze(x,2,0,0);
-->plot2d(x,y)
```

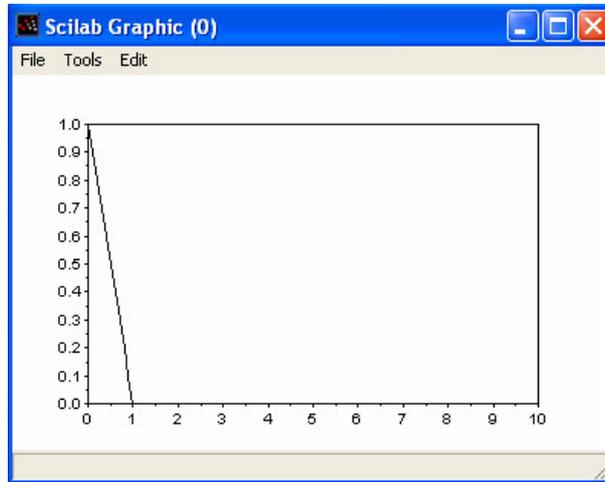

*Fig 5. An example of using the trapeze function*

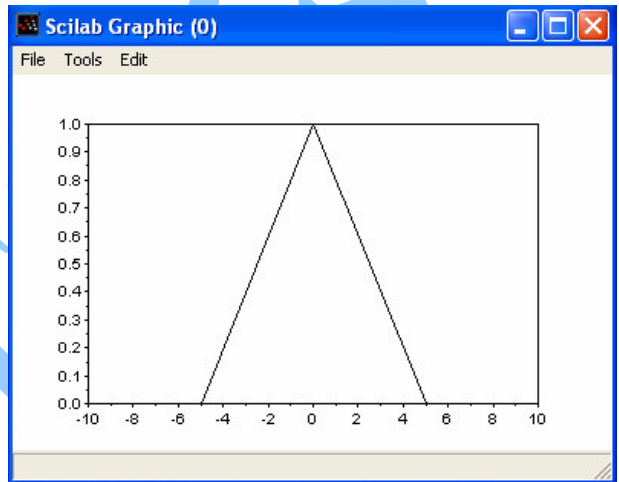

*Fig. 6. An example of creating a belonging function
of the triangular type using trapeze function*

Another example of using *trapeze* function:
```
-->x=(0:1:10)';
-->[y]=trapeze(x,2,2,2);
-->plot2d(x,y)
```

112

leading to the result from fig. 7.

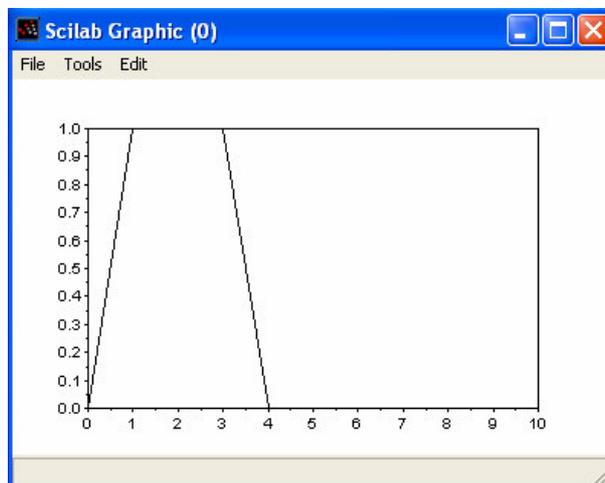

*Fig. 7. Example of using trapeze function*